\documentclass[a4paper,12pt]{article}
\usepackage[T2A]{fontenc}
\usepackage[cp1251]{inputenc}
\usepackage{amsmath,amsthm,amssymb}
\usepackage{graphics}
\usepackage{graphicx}
\begin{document}
\title{Conductance of non-ballistic point contacts in hybrid systems "normal metal/superconductor" Cu/Mo-C and
Cu/LaOFFeAs}
\author{Yu. N. Chiang (1), M. O. Dzyuba (1 and 3), O. G. Shevchenko (1),\\
and A. N. Vasiliev (2)\\
 ((1) B. Verkin Institute for Low Temperature Physics and Engineering,\\
 National Academy of Sciences of Ukraine, Kharkov 61103, Ukraine\\
 (2) Lomonosov Moscow State University,\\
GSP-1, Leninskie Gory, Moscow 119991, Russian Federation,\\
(3) International Laboratory of High Magnetic Fields and Low Temperatures,\\
 53-421 Wroclaw, Poland)\\
 \emph{E-mail:chiang@ilt.kharkov.ua} }
\date{}
\maketitle
\begin{abstract}
We consider the shape of the curves of "Andreev"\ conductance of non-ballistic point-contact NS
heterosystems depending on the bias voltage at the contact. The obtained shape of those curves is
caused by the contribution from the mechanism of coherent scattering by impurities which doubles the
scattering cross section. The behavior of generalized and differential conductance is compared for
ballistic and non-ballistic transport regimes. The criteria are considered allowing one to
discriminate between those regimes with the corresponding conduction curves similar in appearance. The
analysis is extended to the case of non-ballistic transport in NS point contacts with exotic
superconductors, molybdenum carbide Mo-C and oxypnictide La[O$_{1-x}$F$_{x}$]FeAs from a group of
iron-based superconductors.
\end{abstract}
\section{Introduction}
Studies of the physical phenomena underlying the process of percolation flow of charge through the
interfaces of conductive heterosystems are always relevant, because the individual characteristics of
the interfaces such as both quality and geometry, distinguished by diversity, initiate the
consideration of various scenarios of electron scattering at the interfaces. This is especially
true of the interfaces formed by the contacts of normal metals (N) including magnetic ones (F), with
superconductors (S). At present, the investigations of scattering mechanisms in these interfaces are
also stimulated by the emergence of new superconductors based on multicomponent systems with complex
mechanisms of electron correlation [1].

Single charge transfer through the obstacles between the metals in the form of interfaces, overcome by
tunnelling, is not only impeded because of the potential barriers, but also, in the case where one of
the metals is a superconductor, is generally prohibited due to the absence of free states in the
latter for the quasiparticles with energy less than the superconducting energy gap ($\Delta$). The
laws of conservation, however, eliminate this prohibition generating a hole excitation (\emph {h}) in
N - metal in the  direction opposite to that of the velocity of the electron (\emph{e}), incident on
the interface from the N - metal side. The hole has the same energy and momentum as those of the
incident electron, but opposite spin direction. The process is known as Andreev retroreflection
[2]. (Spin orientation of the  quasiparticles in the Andreev \emph{e-h} hybrid which corresponds to a
singlet state, is particularly significant for FS systems, since the magnetic metals, unlike
non-magnetic ones, do not possess symmetry with respect to spin rotation.) Due to this fact, another
electron at the N side is involved in the process of charge transfer through the interface, so that a
part of quasiparticle current is allowed to be directly converted into a supercurrent - the current of
Cooper pairs on the side of a superconductor.
\begin{figure}[htb]
\begin{center}
  \includegraphics[width=10cm]{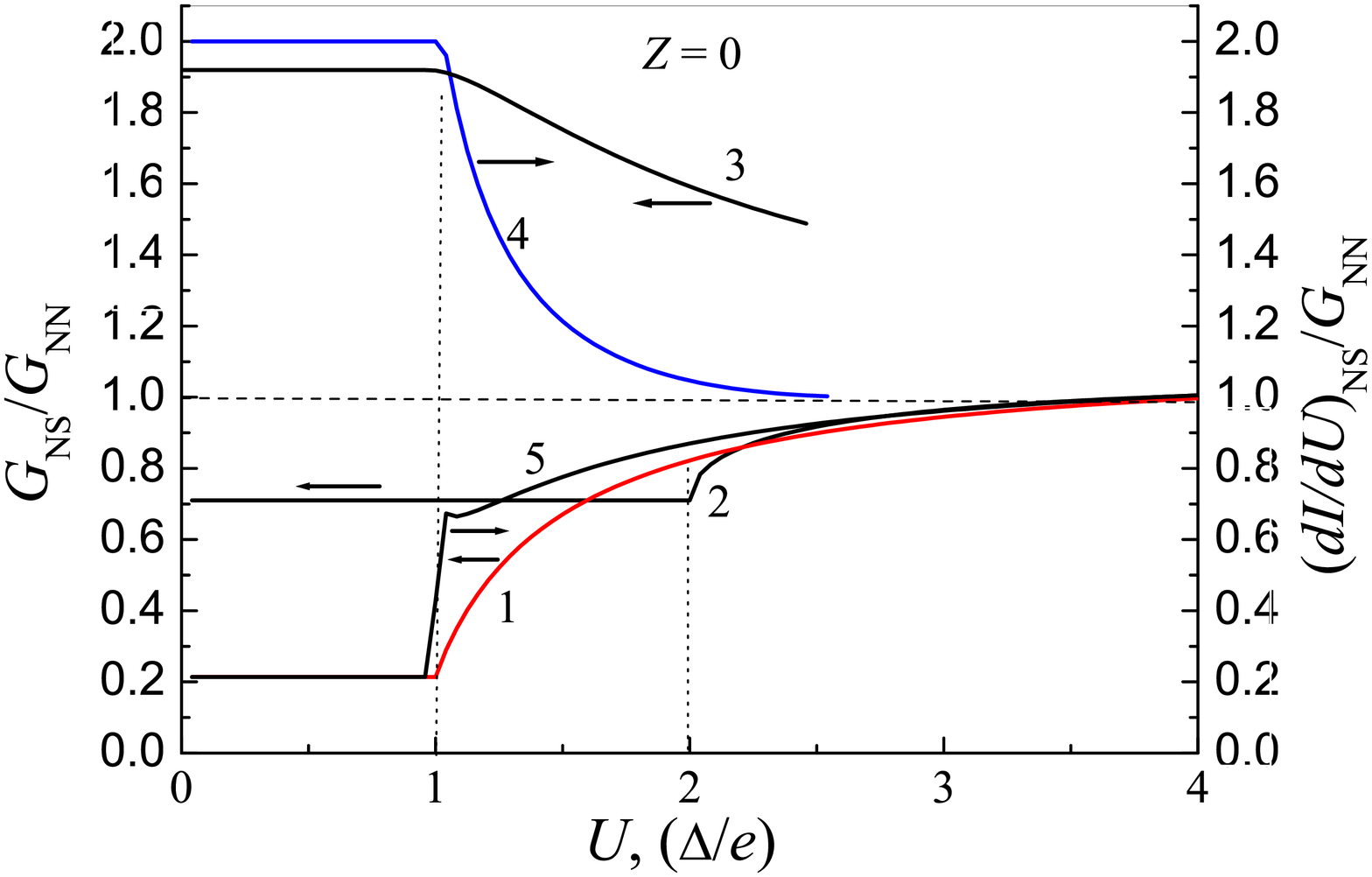}\\
  \caption{Normalized generalized conductance of non-ballistic NS contacts with a barrier-free
  interface (\emph {z} = 0) calculated by Eq. (8): \emph{1} and \emph{2} -
  with $L^{\rm N}/l_{el}^{\rm N} = 1 ~ \mbox {and} ~ L^{\rm N}/l_{el}^{\rm N} = 2$,
  respectively. \emph{3} - generalized conductance of a ballistic  contact; \emph{4} -
  differential conductance of a ballistic contact with the barrier-free interface in
  the BTK model [3]; \emph{5} - differential conductance of a non-ballistic contact
  [Eqs. (8) and (10)]. $T=0$.}
\label{1}
\end{center}
\end{figure}
It follows that at the transition of one of the metals of a bimetallic system from the normal to
superconducting state, this quantum effect can be identified by a change (decrease) in the voltage
\emph{U} when measured in constant-current regime, or by a change (increase) in the current \emph{I}
in constant-voltage regime, if the system has conductance of a finite value. In the limit of NS
bisystem consisting of nondissipative metals with the mean free path of electrons $l_{el}^{\rm N}
\rightarrow \infty$ (the BTK model [3]), the role of this conductance will be played by the only
conductance in the system - that of the interface, $G_{if}$. It will, therefore, concentrate all the
voltage \emph {U} applied to the system, so that $G_{if}=I/U$ (such a quantity we call the generalized
conductance, in contrast to the differential one, $G'(U)=dI/dU$). In this limiting case, the role of
Andreev reflection is reduced to a twofold change in $G_{if}$ at the ${\rm NN} \leftrightarrows {\rm
NS}$ transitions of the system. At energies $eU;\ k_{\rm B}T<\Delta$ (\emph{T} is the temperature),
such change will manifest itself in the form of a \emph{maximum} in conductance curves (see curve
\emph{3} in Fig. 1) if there are no additional sources of scattering at the interface (ie, in case of
barrier-free interface, $z = 0$, in the terminology of Ref. [3]). Note that in the ballistic
approximation, $l_{el}^{\rm N} \rightarrow \infty$, the generalized and differential conductances of a
barrier-free interface are qualitatively of the same shape (compare curves \emph{3} and \emph{4} in
Fig. 1).

In the ballistic transport regime, another form of $G(U)$ curves is also possible when $eU;\ k_{\rm
B}T<\Delta$, namely, the shape of the curve with a \emph{dip} corresponding to a decrease in the
interface conductance with decreasing the transparency of the interface due to existence of a barrier
of finite height, $z \neq 0$, which reduces the probability of Andreev reflection. Thus, in the
ballistic transport model, the only possible shapes of the normalized curves $G(U)/G_{\rm NN}$ (here,
$G_{\rm NN}$ is the generalized conductance in the NN state of the system) in the energy range $eU<
\Delta$, similar to the form of $G'(U)/G_{\rm NN}$ curves in the same interval, are those with a
maximum ($G_{if(\rm NS)}R_{\rm NN}=2,\ z=0$) or with a dip ($0<G_{if(\rm NS)}R_{\rm NN}<2;\ z \neq 0$)
[3].

However, as will be seen later, the aforementioned shape of the curves $G(U)/G_{\rm NN}$ with a dip is
not a prerogative of the ballistic model, as is often assumed. The observation of curves of the
similar shape in special circumstances, which do not produce any justification for using ballistic
conception, indicates the same. The measurement of current-voltage characteristics of the samples with
a point-contact geometry (point contacts) must be considered as such special conditions. With high
probability we can assume that in the majority of point contacts, prepared in different ways, the
ballistic transport conditions are not realized. In this connection, the real curves of point-contact
normalized conductance, although visually resembling the curves for the interface conductance in the
ballistic model, can have completely different nature and lead to conclusions radically different from
those based on the ballistic analysis.
\begin{figure}[htb]
\begin{center}
  \includegraphics[width=10cm]{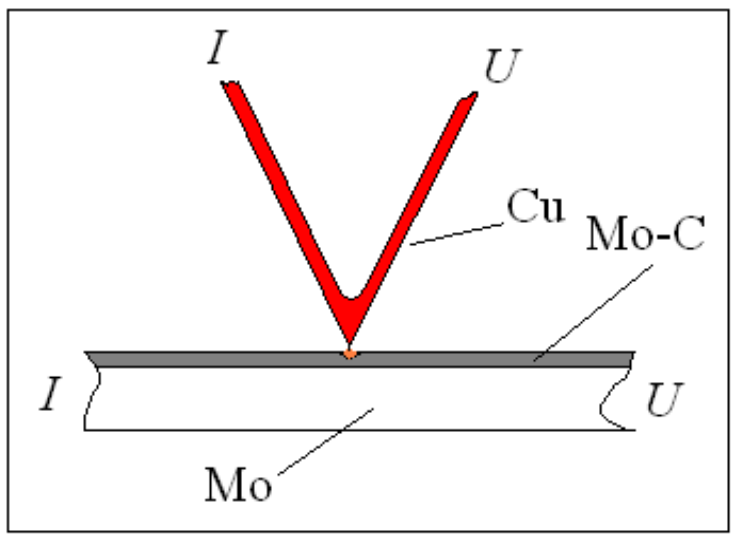}\\
  \caption{Schematic view of measurements of point-contact electrical characteristics.}\label{2}
  \end{center}
\end{figure}
\section {Non-ballistic NS point contact}
The identification of the experimental $G(U)$ curves with the curves of the same shape in the
ballistic NS model means automatically that the voltage measured across a real contact is taken
entirely related to the NS interface and the interface is considered the only possible source of
resistance in the system. In the absence of scattering centers for electrons, the resistance of the
interface, $R_{if}$, is non-zero due to the effect of "bottleneck"\ for incompressible media, thus
depending only on the cross section of the interface $\mathcal{A} \sim d^{2}$ (here, \emph{d} is the
transverse size of the interface). Yu Sharvin was the first to attract attention to this fact [4]:
\begin{equation}\label{1}
R_{if}=(G_{if})^{-1}=(p_{\rm F}/e^{2}n)/d^{2} \equiv 3[2N(0)e^{2}v_{\rm F}]^{-1}/ \mathcal{A}.
\end{equation}
Here, $e,\ p_{\rm F},\ v_{\rm F},\ n,\ \mbox{and}\ N(0)$ are the charge, the Fermi momentum and
velocity, the concentration and density of states of free electrons per spin, respectively.

For N metals Cu, Al, Au, and Ag commonly used in the contacts we have
\begin{equation}\label{2}
R_{if} \approx 5 \cdot 10^{-11}/ \mathcal{A} ~ [\rm Ohm],
\end{equation}
if $\mathcal{A}$ is expressed in cm$^{2}$. Eq. (2) is a criterion for using ballistic approximation
while considering a mode of transport in real point contacts.

As noted above, both geometrical and physical conditions of measurement of the contact conductance in
real experiments cannot meet those of ballistic transport regime. Indeed, a typical longitudinal size
of the contacts, $L_{c}$, defined by the distance between measuring (potential) probes, is of the
order of $10\ \mu$m or more. This is due to the fact that the geometry of point contacts does not
allow one to measure the interface region by a standard four-probe technique, since it appears to be
either three-contact, including significant resistance of a probe (usually), or practically
two-contact, as is shown in Fig. 2. Because of this, the potential lines, which determine the voltage
measured across the probes, in fact, are always on both sides of the interface. They are moved into
the depth of both N and S layers at distances comparable to or larger than $l_{el}^{\rm N}$. As a
result, the actual size of the contact is at least an order of magnitude greater than the thickness of
the interface (for example, the thickness of oxide layers, $t \approx 10 \div 100$ \AA ~ [5]).

In addition, noting that
\begin{equation}\label{3}
(\rho l_{el})^{N} = 3/2N(0)e^{2}v_{\rm F} = R_{if} \mathcal{A}
\end{equation}
($\rho$ is the resistivity), we find that the condition $(R_{c}/R_{if})\geq 1$, under which we cannot
assume that the voltage is applied only to the interface, is achieved at $L_{c}\geq l_{el}^{\rm N}$
for any area of the interface, as this area and N - side cross section are the same, at least within
the limits of the distance from the interface of the order of $l_{el}^{\rm N}$. Hence, an important
conclusion follows: The value of the contact resistance which, on the contrary, depends on the cross
section, cannot be a criterion for the ballistic nature of transport in the contact.

We illustrate this with two examples. Typical magnitude of the resistivity $\rho$ of copper or gold
wires, $\sim 0.1$ mm in diameter, which are mostly used as N - side material (probe) in NS contacts,
at helium temperatures amounts to $\rho \approx 10^{-7}\ {\rm Ohm \cdot cm}$, which value, according
to Eqs. (2) and (3), should correspond to the mean free path $l_{el}^{\rm N} \sim 1\ \mu$m. This is no
less than an order of magnitude shorter than the above estimated $L_{c}$, that is inconsistent with
the ballistic consideration. At the same time, when the contact area $\mathcal{A}$ is equal to the
cross section of the wire of the specified diameter, the measurement will show the total contact
resistance $R_{L_{c}}$ of order of $3 \div 10 ~ \mu$Ohm.

On the contrary, when using the N probe with a working tip size $d \sim 10 \div 100$\ \AA,~ which
value is much less than $l_{0} \approx 0.02$ mm ($l_{0}$ is the mean free path in the absence of size
effect), $l_{el}^{N}$ is approximately equal to $d\ (\ll l_{0})$ and $\rho \sim \rho_{0} (l_{0} / d)
\ln^{-1} (l_{0} / d)$ [6]. In this case, the condition $(R_{c} /R_{if}) \geq 1$ which reflects the
transition from ballistic to diffusive transport will be satisfied in contacts with the resistances of
the order of 10 Ohm or greater at the same liquid-helium temperatures. Clearly, in high-resistance
contacts with the resistance of about 100 Ohm it is unacceptable to estimate the interface area (or
the size of the probe tip) in accordance with Eq. (3), as is often done, since, under $d \ll l_ {0}$,
Eq. (3)  underestimates the size of contact area by a factor of $\rho /\rho_{0}$ (by several orders of
magnitude), thereby creating the illusion of ballistic transport. Thus, the condition $L_{c}\geq
l_{el}^{\rm N}$, inconsistent with the ballistic approximation, appears to be satisfied in most real
point contacts regardless of their total resistance. This means that in real contacts, potential bias
is distributed along the whole length of the contact rather than concentrating at the interface, as
required by the ballistic model.

Consider the metal contact in the absence of oxide layers and additional scattering potentials at the
interface (\emph{z} = 0). In accordance with the additivity rule of resistances (Matthiessen's rule),
generalized contact conductance $G_{\Sigma}$ can be expressed as $G_{\Sigma} = [G_{if}^{-1} + G_{\rm
N}^{-1}]^{-1}$ where $G_{\rm N}$ is the conductance of the normal region before the interface with
cross section $\mathcal A$ and $G_{if} = (R_{t })^{-1}$ is the conductance of the barrier-free
interface. We write the expressions for the current on the N side in the NN and NS states of the
interface. Since, as noted above, $L_{c} \gg t$, then $L_{c}$ actually coincides with the length of
the normal part of the contact, $L^{\rm N}$, and in the NN flowing regime the current can be written
as
\begin{eqnarray}\label{4}
{I_{\rm NN}= j\mathcal{A} = \sigma_{\Sigma (\rm NN)}E \mathcal{A}= \sigma_{\Sigma
(\rm NN)}(U/L^{\rm N}) \mathcal{A}=}
\nonumber\\
=(2/3)N(0)ev_{\rm F}(1 + \frac {L^{\rm N}}{l_{el}^{\rm N}})^{-1} \mathcal{A} (eU)=R_{if}^{-1}(1 +
\frac{L^{\rm N}}{l_{el}^{\rm N }})^{-1}U,
\end{eqnarray}
where in addition to Eqs. (1) and (3) we used $\sigma_{\Sigma (\rm NN)}=G_{\Sigma (\rm NN)}L^{\rm N}/
\mathcal{A}$ and $G_{\Sigma (\rm NN)}=G_{if (\rm NN)}[1 + L^{\rm N}/ l_{el}^{\rm N}]^{-1},\ G_{if (\rm
NN)}=(R_{if})^{-1}$.

Similarly, in the NS transport mode ($T<T_{c},\ T_{c}$ is the critical temperature of the contact S -
side), the current is determined from the condition
\begin{equation}\label{5}
I_{\rm NS}= \sigma_{\Sigma (\rm NS)}(U/L^{\rm N}) \mathcal{A} = G_{\Sigma (NS)}U=G_{if (\rm NS)}(1 +
G_{if (\rm NS)}/G_{\rm N(\rm NS)})^{-1}U.
\end {equation}
Here, the actual interface conductance $G_{if (\rm NS)}$ calculated by the BTK method [3], which takes
into account the jump of the distribution function at the interface and the probability of Andreev
reflection, can be expressed as follows:
\begin{equation}\label{6}
G_{if(\rm NS)}=I_{\rm NS}/U_{\rm NS}=(2/3)N(0)ev_{\rm F} \mathcal{A} J(eU_{\rm NS})/U_{\rm NS} =
R_{if}^{-1}J(eU_{\rm NS})/eU_{\rm NS},
\end{equation}
where
\begin{equation*}
J(eU_{\rm NS})= \int_{-\infty}^{\infty}[f_{0}(E-eU_{\rm NS})-f_{0}(E)][2A+C+D]d(eU).
\end{equation*}
Here, $f_{0}(E)\ \mbox{and}\ f_{0}(E-eU_{\rm NS})$ are the distribution functions at S and N
boundaries of the interface, respectively; $E=k_{\rm B}T$; $A,\ C,\ {\rm and}\ D$ are the
probabilities of Andreev reflection and transitions without and with changing the sign of the wave
vectors of electron states, respectively; $U_{\rm NS}=UR_{if}/R_{N} \cong Ul_{el}^{\rm N}/ L^{\rm N}$
is the potential jump at the NS boundary that defines the jump of the distribution function. Taking
into account that $G_{\rm N(NS)}=R_{if}^{-1}l_{el}^{\rm N}(A)/L^{\rm N}$, where $l_{el}^{\rm N}(A)$ is
the mean free path of electrons in a dissipative normal region, which may depend on the probability of
Andreev reflection (see below), we get Eq. (5) in its final form
\begin{equation}\label{7}
I_{\rm NS}= \{ R_{t}[(J(eU_{\rm NS})/eU_{\rm NS})^{-1} + \frac{L^{\rm N}}{l_{el}^{\rm N}(A)}] \}^{-1}
U.
\end{equation}

An important conclusion follows from this expression for the current in a non-ballistic NS contact:
The value of the bias voltage \emph{U}, at which the transition between the regimes ${\rm NS}
\rightleftarrows {\rm NN}$ should be observed at a given temperature $T<T_{c}$, must be $l_{el}^{\rm
N}/L^{\rm N}$ times higher than the voltage $U_{\rm NS}$ at the interface corresponding to the
superconducting gap energy $\Delta(T)$ of the S side of the contact. Apparently, this is observed in
the majority of point-contact current-voltage characteristics (see, eg, Refs. [7 - 9]). Dividing Eq.
(7) by Eq. (4) we obtain an expression for the normalized conductance of a non-ballistic NS hybrid
system with an ideal (\emph{z}=0) interface:
\begin{equation}\label{8}
\frac{I_{\rm NS}}{I_{\rm NN}} = \frac{G_{\rm NS}}{G_{\rm NN}} = \frac{1+L^{\rm N}/l_{el}^{\rm
N}}{[J(eU_{\rm NS})/eU_{\rm NS}]^{-1}+ L^{\rm N}/[l_{el}^{\rm N}(A)]_{\rm NS}}
\end{equation}

At helium temperatures, under conditions of dominating elastic scattering of electrons by impurities,
the N part of the system is a region of coherent scattering of electrons within a distance $\xi_{\rm
N} \geq \hbar v_{\rm F}/k_{\rm B}T \sim 1\ \mu$m from the NS interface. Hence, $\xi_{\rm N} \geq
l_{el}^{\rm N}$. Under these conditions, the elastic scattering at the same impurity of the electron
and coherently conjugated Andreev hole leads to a twofold increase in the effective scattering cross
section, $\Theta$, at that impurity [10], in particular, to a double \emph{decrease} in the
conductance $G_{\rm N(NS)}$ of the next-to-interface N layer as a whole, $l_{el}^{\rm N}$ in thickness
[11 - 13]. Accordingly, the dependence of mean free path of electrons on the N side of the NS
interface on the probability of Andreev reflection can be expressed as follows [10]:
\begin{equation*}\label{}
[l_{el}^{\rm N}(A)]_{\rm NS}=(\alpha \Theta_{\rm NS})^{-1}=[\alpha \Theta_{\rm N}(1+\gamma A)]^{-1} =
L_{el}^{\rm N}[(1+\gamma A)]^{-1}
\end{equation*}
($\gamma \propto l_{el}^{\rm N}/L^{\rm N}$ in order of magnitude [11]). With this in mind, we note the
main features of the curves $G_{\rm NS}(U)/G_{\rm NN}$ for non-ballistic contacts with barrier-free
($z = 0$) interface in the range of bias voltages $0<U<(L^{\rm N}/l_{el}^{\rm N}) \Delta (T)/e ~
\mbox{when} ~ T \ll T_{c}$. In these circumstances, $J(eU_{\rm NS})/eU_{\rm NS}=(2A)^{-1},\ A = 1\
(C,\ D = 0)$ and
\begin{equation}\label{9}
G_{\rm NS}/G_{\rm NN} = (1 + L^{\rm N}/l_{el}^{\rm N})[\frac{1}{2A}+(L^{\rm N}/l_{el}^{\rm N}) (1 +
\gamma A)]^{-1}.
\end{equation}
Comparison of the terms in square brackets shows that at \emph{z} = 0, in the range of values of the
bias voltage, when the probability of Andreev reflection is equal to 1, twofold increase in the
resistance of non-ballistic contacts due to an increase in the scattering cross section of coherent
quasiparticles by impurities dominates over twofold decrease in the intrinsic resistive contribution
from the interface ($[1+ \gamma A]>[2A]^{-1}$). In particular, in the limit $L^{\rm N} \gg l_{el}^{\rm
N}$, we get $G_{\rm NS}(U)/G_{\rm NN}] \rightarrow 1$.

Fig. 1 shows generalized normalized conductance of a non-ballistic metallic NS point contact with a
barrier-free interface (\emph{z} = 0) as a function of bias voltage at the contact. The dependencies
were calculated numerically by Eq. (8) for two values of the ratio $L^{\rm N}/l_{el}^{\rm N}$, 1
(curve \emph{1}) and 2 (curve \emph{2}), in the approximation $T = 0$. Also shown are the same
conductance obtained by calculation in the ballistic approximation (curve \emph{3}) and the
corresponding differential conductance (curve \emph{4}). When calculating Eq. (8), the function
$A(eU)$ which depends on the ratio of coherent factors from the Bogolyubov equations, was taken from
Ref. [3].

Comparison of the curves \emph{1 - 3} presented in the Figure for the generalized normalized
conductance $G^{\rm NS}(U)/G^{\rm NN}$ shows at least two major differences between ballistic (curve
\emph{3}) and non-ballistic (curves \emph{1, 2}) contacts. First, the value of $U$ at which the
behavior of non-ballistic conductance at the NN $\rightleftarrows$ NS transitions of the system
changes radically, depends on the ratio $L^{\rm N}/l_{el}^{\rm N}$. Second, the asymptotics of $G^{\rm
NS}(U)/G^{\rm NN}$ are of the opposite nature, both in $eU<\Delta$ and in $eU>\Delta$ regions.
Usually, they analyze the differential conductance, as more expressive, rather than the generalized
one. In Fig. 1, curve \emph{5}, we show the behavior of the normalized differential conductance of the
non-ballistic contact
\begin{equation}\label{10}
\frac{(dI/dU)}{G_{NN}}=U \frac{d(G_{\rm NS}(U)/G_{\rm NN})}{dU}+\frac{G_{\rm NS}(U)}{G_{\rm NN}}
\end{equation}
for the same ratio $L^{\rm N}/l_{el}^{\rm N}=1$ that corresponds to the generalized conductance
depicted by curve \emph{1}. It is seen that the main features of the asymptotic behavior in both cases
are common.
\begin{figure}\begin{center}
  \includegraphics[width=10cm]{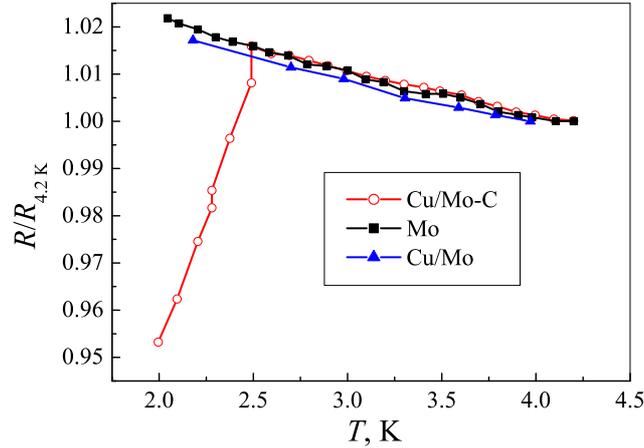}\\
  \caption{Temperature dependencies of resistance for: NS point contact Cu/Mo-C
  without oxide layer - $\bigcirc$; NN point contact Cu/Mo - $\blacktriangle$;
  non-carbonized Mo - $\blacksquare$.}\label{3}
\end{center}
\end{figure}
Thus, the asymptotic behavior of the conductance of non-ballistic contacts appears \emph{formally
similar} to the behavior of $G'_{if}$ of a ballistic contact with the barriers $z \neq 0$ at the
interface (see Ref. [3]), which fact is apparently the reason for the application of a ballistic model
(often unjustified) while analyzing the curves $G(U)$. Since in this case details of the curves assume
a meaning which does not correspond to reality, it can easily lead to wrong conclusions.

Special note is expedient to make in regard to non-ballistic tunnel junctions of hold-down type, with
the oxide interface undestroyed, which have semiconductor conductivity as a rule. The resistance of
these interfaces can be the greatest in the contact, even in the NN state. In this case, in this
state, all the bias voltage will drop at the interface, as a ballistic model assumes. However, since
the thickness of the oxide layers \emph{t} is so small that $t \ll \xi_{\rm N}$ (see above), then, at
the transition from the NN to NS state, due to the proximity effect, the resistance of the oxide layer
vanishes, having experienced the superconducting transition in the parameter \emph{eU}. As a result, a
part of the curve of the generalized conductance $G^{\rm NS}(U) /G^{\rm NN}$ in some interval of $eU ~
\mbox{close to}~ \Delta$ at first takes the shape similar to that of the curves \emph{3} or \emph{4}
in Fig. 1. However, at $eU<\Delta$, the potentials will be redistributed, thus transforming transport
to the regime where the conductance behaves in accordance with Eq. (8). In high-resistance contacts,
the scale of this part of the curve is usually small against that part of the incomparably larger
scale, which is associated with the manifestation of the above proximity effect, so that the overall
shape of the curve may easily be taken for a behavior of the ballistic conductance to which it has
nothing to do.

In this paper, we investigate the behavior of the conductance of non-ballistic point-contact hybrid NS
systems with unconventional superconductors, Cu (N) / molybdenum carbide Mo-C (S) and Cu (N) /
oxypnictide La[O$_{1-x}$F$_{x}$]FeAs (S), $x \approx 0.15$, in the Andreev reflection mode.
\begin{figure}[htb]
\begin{center}
\includegraphics[width=10cm]{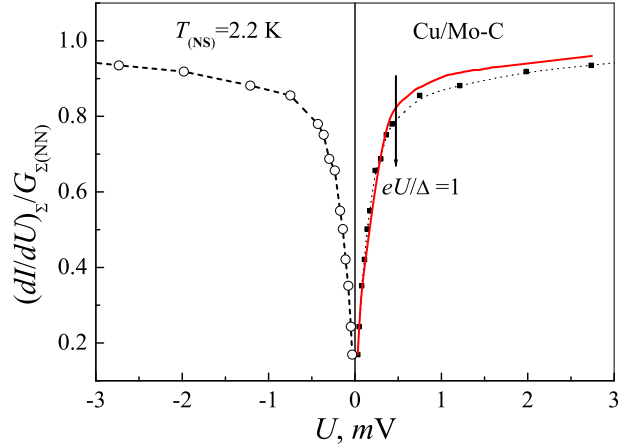}\\
\caption{Dependence of the normalized differential conductance of a non-ballistic NS point contact
Cu/Mo-C with the oxide layer destroyed on the bias voltage across the contact. $T = 2.2$ K. Solid line
is a \emph{fit} in accordance with Eq. (10) from the text.}\label{4}
\end{center}
\end{figure}
\section{Experiment}
25 hybrid contacts have been studied. In Fig. 2, the location of the probes for measuring the
characteristics of contacts, including current-voltage ones, is shown. All measurements were performed
at a constant dc current. It means that the internal resistance of the current source (this includes
regulating resistances) was always greater than the resistance of the contacts. We specifically
stipulate that as, while studying high-resistance tunnel NS contacts with semiconducting properties
(eg, about 100 Ohm and above), at NN $\rightleftarrows$ NS transitions, in addition to those features
mentioned earlier, the probability of switching preset regimes is not excluded yet, when using a
current source with rather low resistance. This can also distort the meaning of the results obtained.
\begin{figure}[tb]
\begin{center}
\includegraphics[width=10cm]{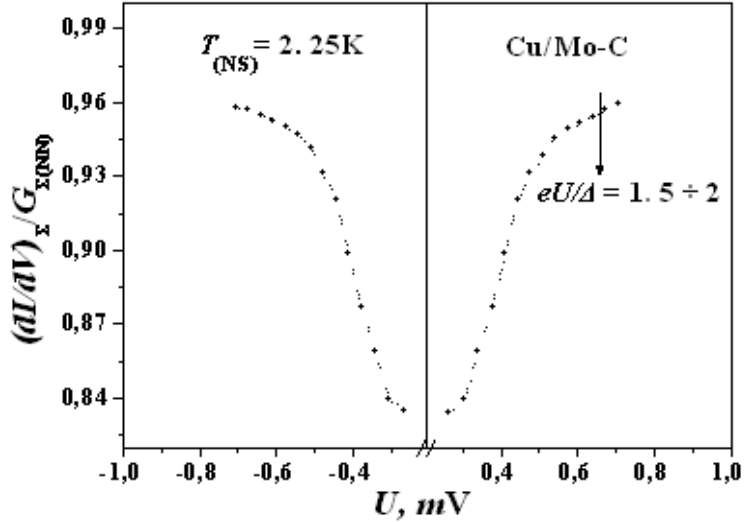}\\
\caption{Dependence of the normalized differential conductance of a non-ballistic NS point contact
Cu/Mo-C with the oxide layer destroyed on the bias voltage across the contact. $T = 2.25$ K.}\label{5}
\end{center}
\end{figure}

Contacts were made in two ways - by the spark erosion method and mechanically. Contacts made by the
first method (Cu/molybdenum carbide Mo-C) have low resistance since the tip touches directly with the
superconductor burning the oxide barrier; besides, the area of melted tip in this case was
considerable. Contacts made by the second method (Cu/Mo-C and Cu/La[O$_{1-x}$F$_{x}$]FeAs) have higher
resistance due to conservation of the oxide layer. The difference in resistance between
low-resistivity contacts and high-resistivity ones were within two - five orders of magnitude (from
$10^{-5}$ Ohm to 1 Ohm). The $G(U)$ curves for low-resistance contacts were studied using a
piсovoltmeter [14], for high-resistance ones - with the help of a standard measuring technique. Since
superconducting carbide layer in Cu/Mo-C samples could be completely destroyed at the point of contact
during preparing them by the spark erosion method, we tested the possibility of maintenance of the NS
regime in the process of measuring IVC by controlling the type of the temperature dependence of the
contact resistance. In this study, we investigated the contacts prepared on the same samples of
molybdenum with carbonized surface layer that were described in Ref. [15]. Contacts
Cu/La[O$_{1-x}$F$_{x}$]FeAs were prepared only by hold-down method. In the NN state, they revealed
semiconducting behavior, as well as hold-down contacts Cu/Mo-C.
\subsection*{NS contacts Cu (N) / Mo-C (S)}
Typical temperature dependencies of the normalized resistance are shown in Fig. 3 for the contact with
an undefeated layer of superconducting molybdenum carbide at the interface Cu/Mo-C (circles) and for
the contact with a destroyed superconducting layer, ie, Cu/Mo (triangles). Squares depict the
resistance of non-carbonized parts of molybdenum measured by a standard four-probe technique. The
measured resistance values were normalized to the corresponding contact resistance at 4.2 K. In
particular, for a low-resistance contact (the curve with circles), it amounts to $R(4.2 {\rm K})=
1.43935 \cdot 10^{-5}$ Ohm. The dependence of resistance on temperature of this kind, which contains a
singularity of the superconducting transition type, indicated the possibility of realizing the
transitions NN $\rightleftarrows$ NS at the contact interface by changing the bias voltage \emph{U} to
a value corresponding to an electron energy of the order of $k_{\rm B}T_{c }$. In our samples of
molybdenum carbide, $T_{c}$ is approximately 2.7 K [15].
\begin{figure}[htb]
\begin{center}
\includegraphics[width=10cm]{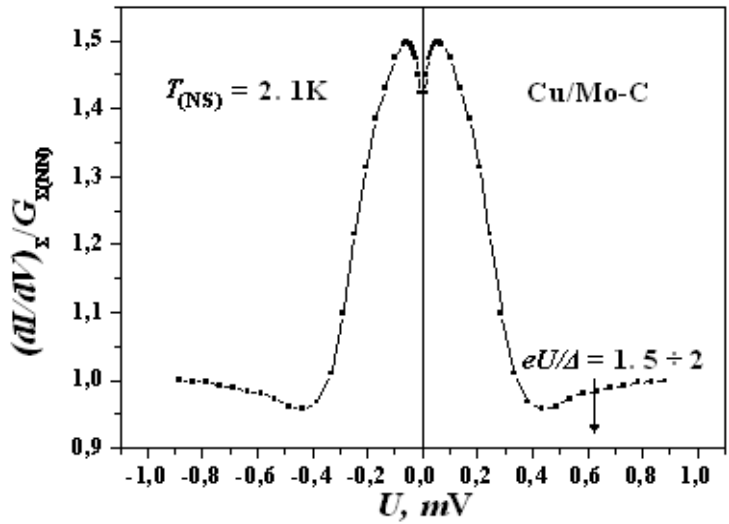}\\
\caption{Dependence of the normalized differential conductance of a non-ballistic hold-down NS point
contact Cu/Mo-C with an undefeated oxide layer on the bias voltage across the contact. $T = 2.1$
K.}\label{6}
\end{center}
\end{figure}
Only the conductance of the contacts which revealed such temperature dependence showed features
associated with the above transitions in the parameter $eU$. Thus, at the transition to the Andreev
reflection regime (NN $\rightarrow$ NS), when $eU / \Delta <1$, the conductance of such contacts
always \emph{reduced} rather than increased, as it could be in the ballistic regime of current flow
(see Fig. 1).

In Figs. 4 and 5, the examples are shown of the behavior of the derivatives of generalized conductance
of the investigated contacts with a burned oxide layer, but an undestroyed superconducting carbide
Mo-C layer, in the range of bias voltage values $eU$ of the order of $\Delta = 1.73 k_{\rm B}T_{c}
\approx 0.47$ mV. Based on the analysis from the previous section, one can assume that the measured
curves describe the behavior of the conductance of a non-ballistic contact with a ratio $L^{\rm
N}/l_{el}^{\rm N}$ of about 1 (Fig. 4) or slightly higher (Fig. 5). When the diameter of the interface
(probe tip diameter) is about 10 $\mu$m, an estimation gives the values of the mean free path of
electrons in the normal contact area and the size of that area of the order of 0.1 $\mu$m.

Figure 6 for a contact Cu/Mo-C and Figure 7 for contacts Cu/La[O$_{1-x}$F$_{x}$]FeAs can serve as an
illustration of the remark made in the previous section concerning the influence of the proximity
effect on the shape of the conductance curves for high-resistance hold-down contacts with an
undestroyed oxide interface. (We will discuss the latter Figure in more detail below.) In these
(always high-resistive) contacts, resistance ratio $R_{c}(4.2\ {\rm K})/R_{c}(300$ K) exceeds 1 which
fact indicates the presence of an oxide layer with non-metallic conductivity at the interface. If $T
<T_{c}$, the contribution from this layer to the contact resistance due to the proximity effect should
be absent, as noted earlier, until the electron energy $eU$ is less than $\Delta$. In this energy
range, the probability of Andreev reflection is close to unity and its contribution to the contact
resistance is maximum (see a minimum in the curves near $U=0$ in Figs. 6 and 7), since in that range,
the mechanism of doubling cross section for scattering of coherent quasiparticles by impurities is
responsible for the decrease in conductance. As the energy $eU$ increases, this resistive
contribution, according to Eqs. (9) or (10) (see curves \emph{1} and \emph{5} in Fig. 1), will
continue to decrease (increasing conductance) until the order parameter at the interface is suppressed
(suppressed proximity effect) and the contribution from a resistive oxide layer becomes to appear, up
to the full suppression at $eU > \Delta$. Note that an indistinguishable contribution from the
boundary resistance on the S side of the contact is not excluded; that contribution is associated with
the charge unbalance of the quasiparticles due to the dispersion of the order parameter at the NS
interface, which behaves similarly [16]. In Fig. 6, corresponded to this are those branches of the
conductance curve that fall to a value close to unity, with increasing $eU$.

In general, one could conclude that the size $L_{c}$ of non-ballistic contacts with the resistance of
no more than 1 Ohm, prepared according to techniques used by us, is of the order of 0.1 $\mu$m for
$(L_{c} / l_{el}^{\rm N}) \approx 1 \div 2$. There exist non-ballistic contacts with higher
resistance, apparently for $(L_{c} / l_{el}^{\rm N}) \geq 10$ [7 - 9].
\begin{figure}[htb]
\begin{center}
\includegraphics[width=10cm]{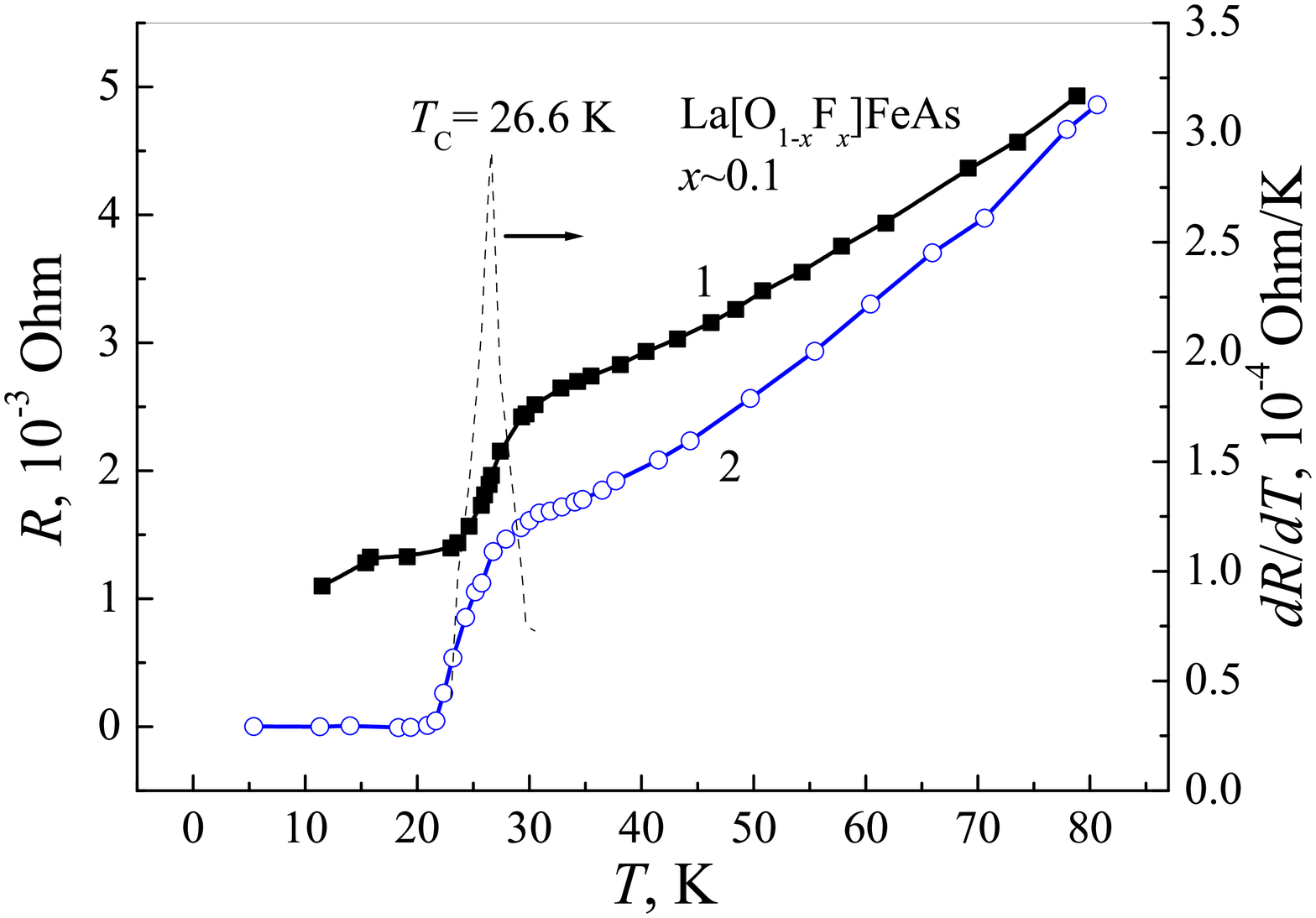}\\
\caption{Superconducting transition in the temperature dependence of resistance for the oxypnictide
samples we used to prepare NS point contacts: \emph{1} - two-probe method of measurement, \emph{2} -
four-probe method of measurement. Dashed curve is the differential of curve \emph{2}.} \label{7}
\end{center}
\end{figure}
\subsection*{NS contacts Cu (N)/ La[O$_{0.85}$F$_{\approx 0.1}$]FeAs (S)}
The emergence of new iron-based superconductors - oxypnictides - caused another burst of interest in
the problem of compatibility between superconductivity and magnetism. Experimental investigations of
the conductance of point-contact NS heterosystems, with oxypnictides as a superconductor, under
conditions of Andreev reflection, seem to be an important trend in the study of this problem. The
Andreev reflection is known to lead, in the absence of magnetic fields, to initiation of coherent
quasiparticles, \emph{e} and \emph{h}, on N side which are related through that reflection at the NS
interface. Their coherent scattering by impurities leads, as noted, to a significant decrease in the
conductance of non-ballistic NS systems due to doubling scattering cross section. Thus, the behavior
of the conductance of NS systems, in particular, with changing the electron energy \emph{eU},
allows one, above all, to ascertain whether the conductance of the system is of Andreev type and,
therefore, to understand, whether the intrinsic magnetism of oxypnictides plays any role in the
manifestation of the effects associated with the superconducting properties.

Figure 7 confirms the existence of the superconducting transition in the temperature dependence of
resistance of two oxypnictide samples we used to prepare NS point contacts, the conductance of which
was then investigated in the regime of Andreev reflection at $T \ll T_{c}$. Temperature dependence of
the resistance of one of the samples was measured by two-probe technique (curve \emph{1}) and of the
second (curve \emph{2}) - by four-probe technique. Indeed, a superconducting feature occurred in the
curves \emph{R(T)} at 26.6 K, as was also observed for oxypnictides La[O$_{1-x}$F$_{x}$]FeAs with $x
\approx 0.15$ [1]. We prepared point contacts on these samples by exerting pressure without puncturing
the oxide layer, as evidenced by a significant contact resistance under normal current flow (about 10
Ohm at $T>T_{c}$) and the ratio $R_{c}(78\ {\rm K})/R_{c}(300\ {\rm K})>1$.
\begin{figure}[htb]
\begin{center}
\includegraphics[width=10cm]{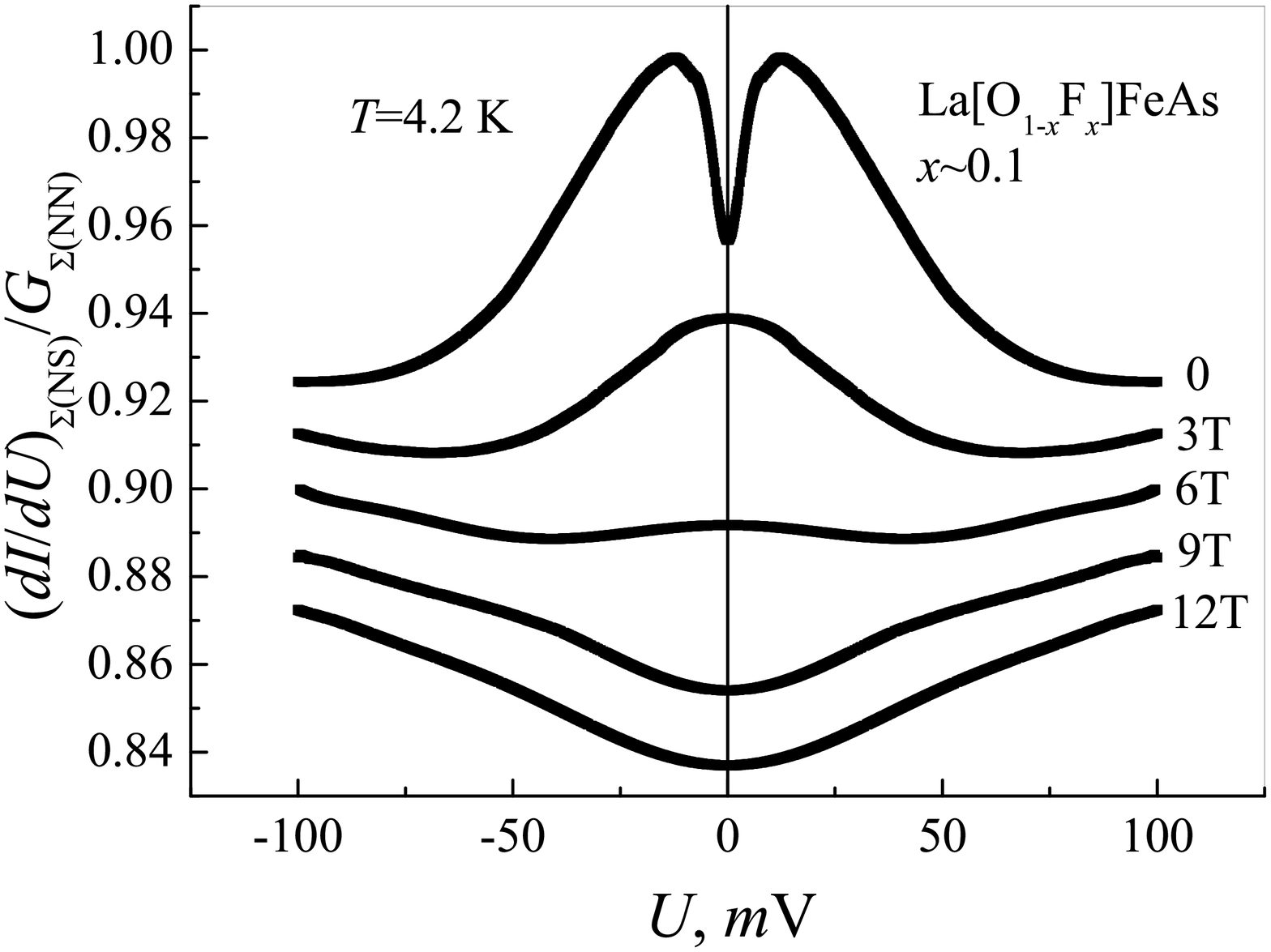}\\
\caption{Dependence of the normalized  differential conductance of a non-ballistic NS point contact Cu
(N)/La[O$_{0.85}$F$_{\approx 0.1}$]FeAs (S) with an undefeated oxide layer on the bias voltage across
the contact in various magnetic fields.}\label{8}
\end{center}
\end{figure}

Typical behavior of the conductance of such contacts as a function of bias voltage across the contact
is shown in Fig. 8 over a wide range of magnetic fields, \emph{B}. We see that the feature that can be
called a "gap feature"\ is observed only in the conductance curve measured in the absence of an
external magnetic field. This means that the intrinsic magnetism of oxypnictides at $B = 0$ does not
prevent either the appearance of superconductivity or other effects associated with it, in particular,
such as Andreev reflection. This is proved, firstly, by the shape of the feature in the curve for
$B=0$, which just corresponds to a decrease in conductance with decreasing energy $eU$, as it should
be in the NS regime of non-ballistic contacts when the cross section for the impurity scattering of
coherent quasiparticles, coupled by Andreev reflection, is doubled. And, secondly, by the low values
of the energy \emph{eU} compared with $\Delta$, for which the effect exists, that is the values
insufficient for suppressing the proximity effect at the interface, and thus, for the manifestation of
a non-zero contribution from the boundary resistance, in particular, the resistive contribution from
the oxide layer. Since under conditions of non-ballistic transport in the contact, a bias voltage is
distributed between the interface and N side of the contact, a position of a gap feature, as shown in
Section 2, depends on the ratio $L^{\rm N}/l_{el}^{\rm N}$ (see Fig. 1). From the shape of the curve
shown in Fig. 9, which is a part of the curve $B = 0$ in Fig. 8 in an enlarged energy scale, it
follows that the value of the gap BCS energy  of the oxypnictide, $\Delta = 1.73 k_{\rm B}T_{c}
\approx 2.5$ mV, \emph{at the interface} is reached at $eU \approx 3 \Delta$ \emph{at the contact}.
Evaluation by Eqs. (8) and (9) shows that the depth of the minimum in the curve at $eU \rightarrow 0$
must be $G_{\rm NS}/G_{\rm NN} \approx 4/4.1 \approx 0.95$ ($G_{\rm NN} = G_{\rm NN} (добавила N) (eU
= 3 \Delta)$), which value, as we see, corresponds to the measured one. The normalized conductance
curves of point-contact NS systems (including those with an oxypnictide as a superconductor) studied
in Ref. [16] can also serve as a good illustration of the discussed features of non-ballistic contact
conductance.

The absence of the above effect in external magnetic fields emphasizes, in addition, that this effect
is a feature of "Andreev"\ conductance in conditions of coherence of \emph{e} and \emph{h}
quasiparticles which takes place only in a zero magnetic field. In a nonzero magnetic field, the
coherence of the quasiparticles is disturbed due to spatial divergence of their trajectories after
Andreev reflection at NS boundary [18]. As can be seen from Fig. 9, when the bias voltage \emph{eU}
exceeds the energy value $3 \Delta$, a non-zero "boundary resistance"\ (the resistance of the oxide
layer) appears which was suppressed before by the proximity effect and is influenced by a magnetic
field. We omit the detailed discussion of this branch of the curves as not relevant to the merit.
However, the impression is that the variation of its shape upon increasing the field is associated
with a decrease in carrier mobility in a semiconductor material. The dependence of this part of the
contact resistance on the magnetic field is shown in Fig. 10, in which $\Delta G^{*}/G(eU = 3 \Delta)
= [G(eU = 100\ {\rm mV}) - G(eU = 3 \Delta)]/G(eU = 3 \Delta)$.
\begin{figure}[htb]
\begin{center}
\includegraphics[width=12cm]{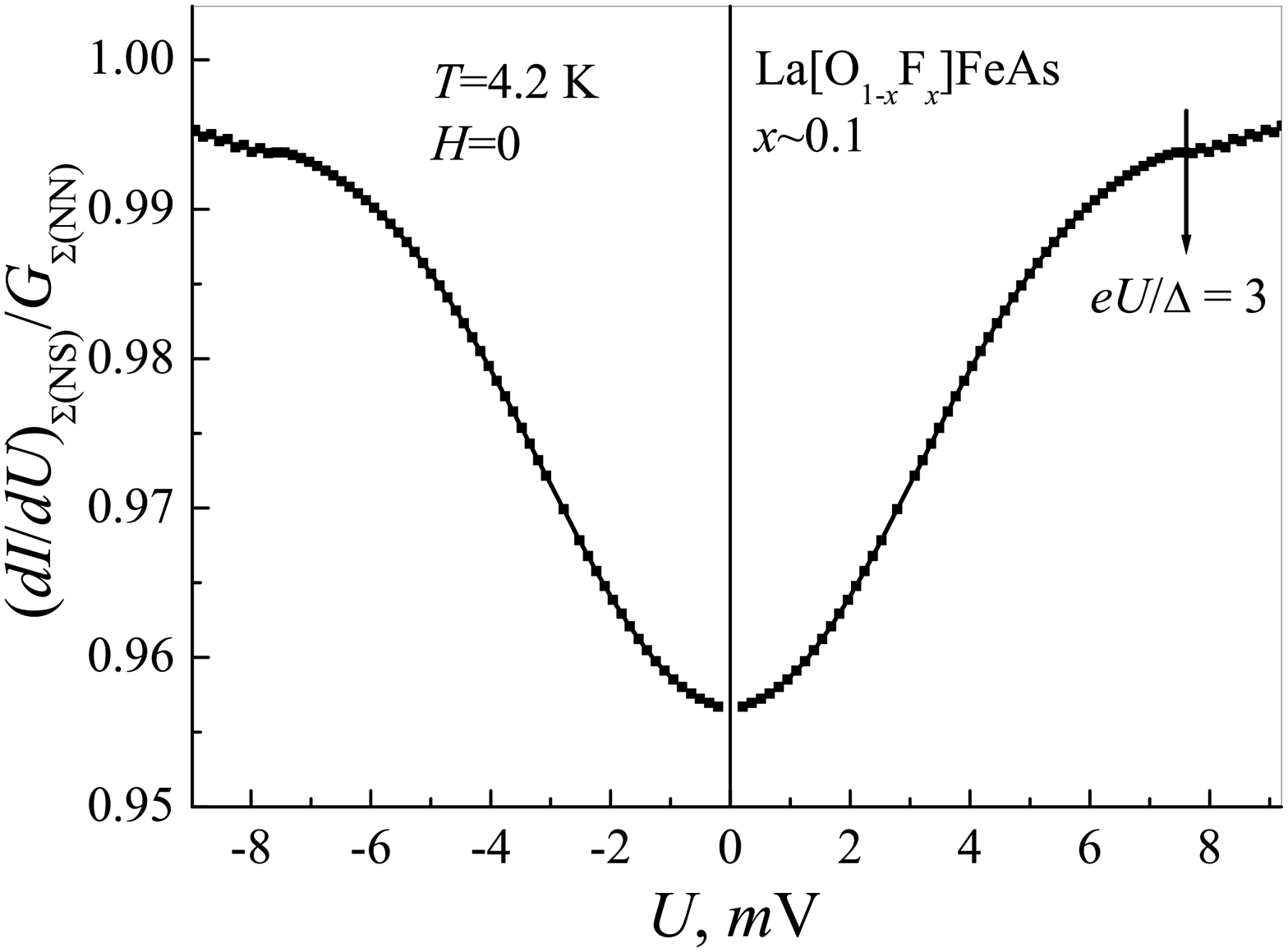}\\
\caption{The increased part of the curve at $B = 0$ from Fig. 8.}\label{9}
\end{center}
\end{figure}
Thus, the experiment shows that the presence of a magnetic element in the structure of an oxypnictide
has no effect on the manifestation of the phenomenon of Andreev reflection in NS systems with
oxypnictides as a superconducting side of the NS interface in the absence of an external magnetic
field. In this connection, one can assume that the mechanisms for magnetic interaction, such as
ferromagnetic fluctuations, and the mechanism for superconducting pairing of charge carriers (as the
results convince, BCS-like) do not overlap in energy in the layered compound La[O$_{1-x}$F$_{x}$]FeAs
with a 10 \% replacement of oxygen by fluorine. It is apparently achieved by a corresponding
fragmentation of the Fermi surface of conduction electrons (nesting), which eliminates the competition
of these mechanisms at the antiferromagnetic long-range order [19, 20]. The suppression of
superconductivity and its attendant effects is observed only in external magnetic fields (in our
experiments, at $B \geq 3$ T).
\begin{figure}\begin{center}
\includegraphics[width=10cm]{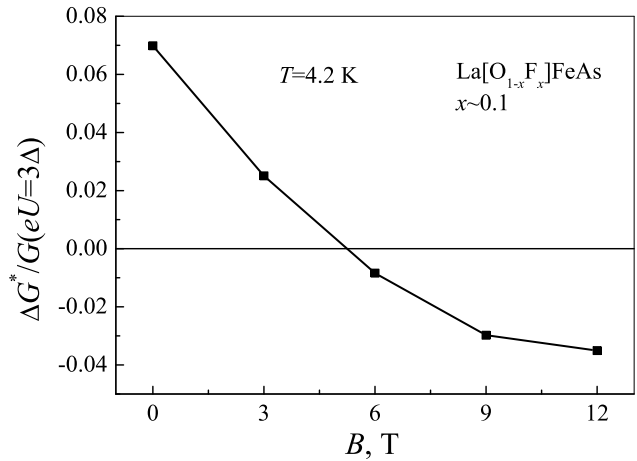}\\
\caption{Dependence of the boundary resistance on a magnetic field in the NN state of a contact at the
suppression of the proximity effect by energy $eU> \Delta$.}\label{10}
\end{center}\end{figure}
\section {Conclusion}
In this paper, we consider the behavior of the conductance of point-contact non-ballistic NS
heterosystems depending on the electron energy $eU$ specified by the bias voltage \emph{U} across the
contact. It is noted that $G(U)$ curves for non-ballistic and ballistic contacts, having a completely
different nature, may, however, coincide in shape, which fact, apparently, is the reason for using,
often unjustified, a ballistic model to analyze those curves for real contacts, which are, in fact,
non-ballistic. It is shown that the resulting shape of the $G(U)$ curves of non-ballistic contacts is
determined by the contribution from the mechanism of coherent scattering by impurities. The higher is
the probability of Andreev reflection, the greater is that contribution, the greatest at the
zero-height barrier at the interface ($z=0,\ A=1$). The same shape of the conductance curve can be
obtained in the ballistic model by fitting the height of the barrier at the interface, thus, by
\emph{suppressing} accordingly the probability of Andreev reflection. Criteria allowing one to
differentiate between the regimes of ballistic and non-ballistic transport are given, at apparent
similarity of the corresponding conductance curves. One of the major indications of the non-ballistic
nature of transport in a contact is the excess of the bias voltage, applied to the contact, above the
voltage corresponding to the energy gap of the used superconductor at which NN $\rightleftarrows$ NS
transition is realized. It manifests itself in the form of peculiarities (kinks) in the conductance
curves $G(U)$. The analysis is extended to the case of non-ballistic transport in NS point contacts
with exotic superconductors, namely, molybdenum carbide, Mo-C, and iron-based oxypnictide,
La[O$_{1-x}$F$_{x}$]FeAs. The experiment shows that the conductance behavior of the real point
contacts of any nature is almost always match that expected for non-ballistic transport regime.
\renewcommand{\refname}{References}

\end{document}